\documentclass[10pt,conference]{IEEEtran}

\IEEEoverridecommandlockouts
\usepackage{cite}
\usepackage{amsmath,amssymb,amsfonts}
\usepackage{algorithmic}
\usepackage{graphicx}
\usepackage{textcomp}
\usepackage{tabularx}
\usepackage{multirow, makecell}
\usepackage{multicol}
\usepackage{xcolor}
\usepackage{siunitx}
\usepackage{booktabs}
\usepackage{subfigure}
\usepackage{hyperref}

\DeclareMathOperator*{\argmax}{argmax}

\def\BibTeX{{\rm B\kern-.05em{\sc i\kern-.025em b}\kern-.08em
    T\kern-.1667em\lower.7ex\hbox{E}\kern-.125emX}}

\begin{document}
\title{Cost-Efficient Multi-Scale Fovea for Semantic-Based Visual Search Attention\\
\thanks{This work was supported by \textit{Fundação para a Ciência e Tecnologia} (FCT) through project (1801P.01460.1.03) LARSYS/ISR BASE 2025-2029 - VISLAB LAB/ISR (DOI: 10.54499/UID/50009/2025). 

João Luzio is supported by the FCT doctoral grant [2024.00683.BD].}
}

\author{
\IEEEauthorblockN{João Luzio}
\IEEEauthorblockA{\textit{Institute for Systems and Robotics} \\
\textit{Instituto Superior Técnico}\\
Lisbon, Portugal \\
joaoluzio14@tecnico.ulisboa.pt}
\and
\IEEEauthorblockN{Alexandre Bernardino}
\IEEEauthorblockA{\textit{Institute for Systems and Robotics} \\
\textit{Instituto Superior Técnico}\\
Lisbon, Portugal \\
alex@isr.tecnico.ulisboa.pt}
\and
\IEEEauthorblockN{Plinio Moreno}
\IEEEauthorblockA{\textit{Institute for Systems and Robotics} \\
\textit{Instituto Superior Técnico}\\
Lisbon, Portugal \\
plinio@isr.tecnico.ulisboa.pt}
}

\maketitle

\begin{abstract}
Semantics are one of the primary sources of top-down preattentive information. Modern deep object detectors excel at extracting such valuable semantic cues from complex visual scenes. However, the size of the visual input to be processed by these detectors can become a bottleneck, particularly in terms of time costs, affecting an artificial attention system's biological plausibility and real-time deployability. Inspired by classical exponential density roll-off topologies, we apply a new artificial foveation module to our novel attention prediction pipeline: the Semantic-based Bayesian Attention (\textit{SemBA}) framework. We aim at reducing detection-related computational costs without compromising visual task accuracy, thereby making \textit{SemBA} more biologically plausible. The proposed multi-scale pyramidal field-of-view retains maximum acuity at an innermost level, around a focal point, while gradually increasing distortion for outer levels to mimic peripheral uncertainty via downsampling. In this work we evaluate the performance of our novel \textit{Multi-Scale Fovea}, incorporated into \textit{SemBA}, on target-present visual search. We also compare it against other artificial foveal systems, and conduct ablation studies with different deep object detection models to assess the impact of the new topology in terms of computational costs. We experimentally demonstrate that including the new \textit{Multi-Scale Fovea} module effectively reduces inherent processing costs while improving \textit{SemBA}'s scanpath prediction accuracy. Remarkably, we show that \textit{SemBA} closely approximates human consistency while retaining the actual human fovea's proportions.
\end{abstract}


\IEEEpeerreviewmaketitle

\section{Introduction} \label{intro}

Human attention is guided by multiple sources of preattentive information \cite{gs6}, such as top-down and bottom-up features, task-related rewards, prior knowledge, scene context, and semantics. Information coming from all these sources is then combined into a single spatial priority map \cite{activevision,trad_sal}, often referred to as the attention map. The relevance of each source is conditioned by the task at hand. On the one hand, bottom-up salient cues and scene syntax may be of utmost relevance for exploratory tasks \cite{predvisualfix}, such as free-viewing. On the other hand, top-down semantics and task-related rewards play a crucial role in goal-directed activities \cite{wolfe}, such as visual search. 

The extraction of information from these sources is also conditioned by the physiology of the human retina. The eye exhibits a higher photoreceptor density around a central focal region \cite{fov} known as the fovea. Within this region, objects are perceived with maximum visual acuity \cite{fov_review} and are therefore easier to identify. As the distance from the fixated point increases, photoreceptor density decreases, leading to progressively blurrier percepts in more peripheral regions. This phenomenon, known as the eccentricity effect \cite{eccentricity}, makes objects located further in the retinal periphery increasingly difficult to recognize, due to the intensifying distortion levels.

\begin{figure}
    \centering
    \includegraphics[width=0.95\linewidth]{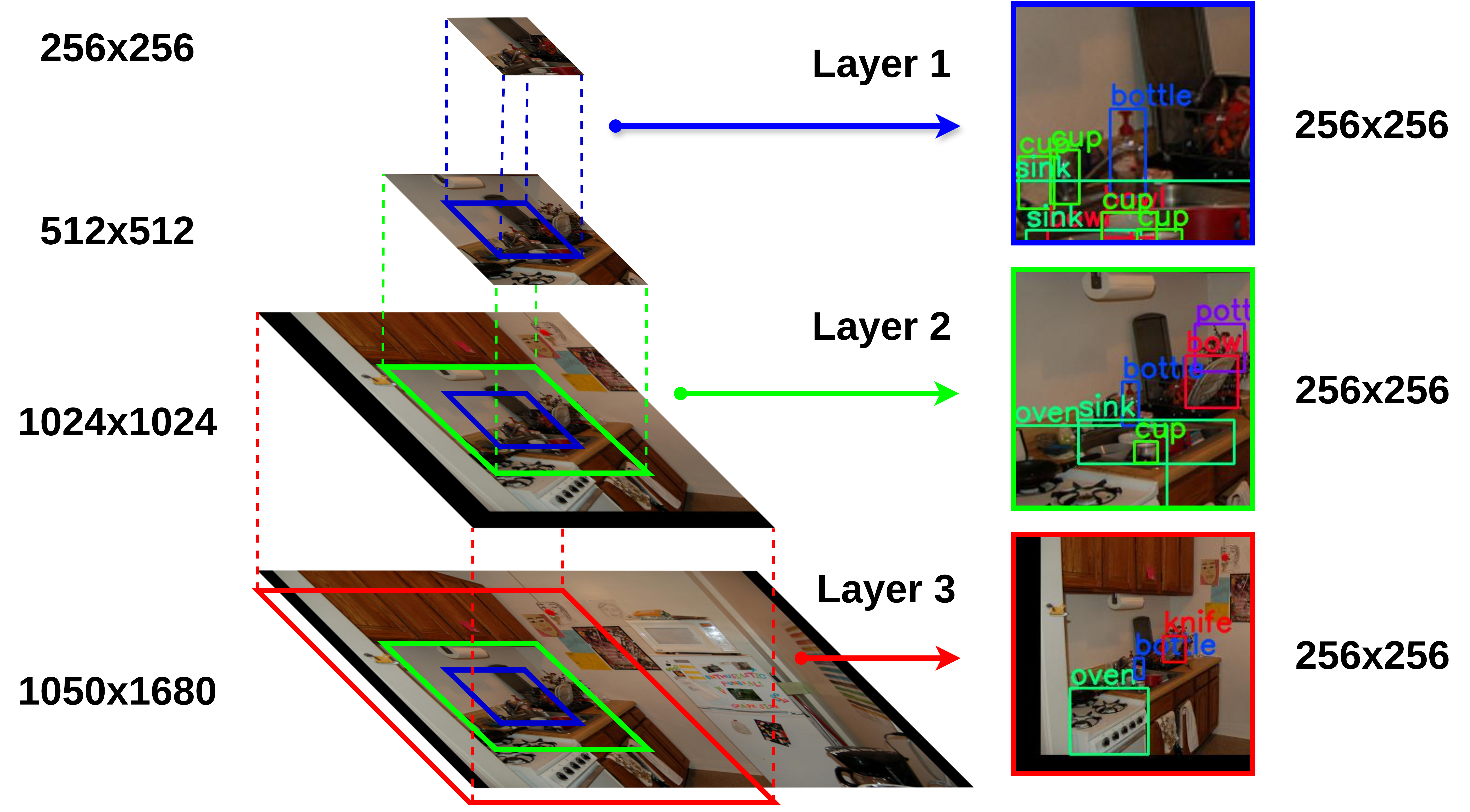}
    \caption{Illustration of our novel \textit{Multi-Scale Fovea} mechanism. The proposed method consists of building a multi-resolution pyramid \cite{image_pyramid}, around a selected focal point, and then downsampling all levels to the size of the innermost layer, to mimic the eccentricity effect \cite{eccentricity}. Object detections from outer levels tend to reflect the uncertainty that derives from such exponential pixel density reduction \cite{od_pyramid}. This technique facilitates the sequential extraction of semantic information \cite{icdl} in a more cost-efficient and biologically plausible manner.}
    \label{fig:cover}
\end{figure}

\begin{figure*}
\centering
\includegraphics[width=0.95\linewidth]{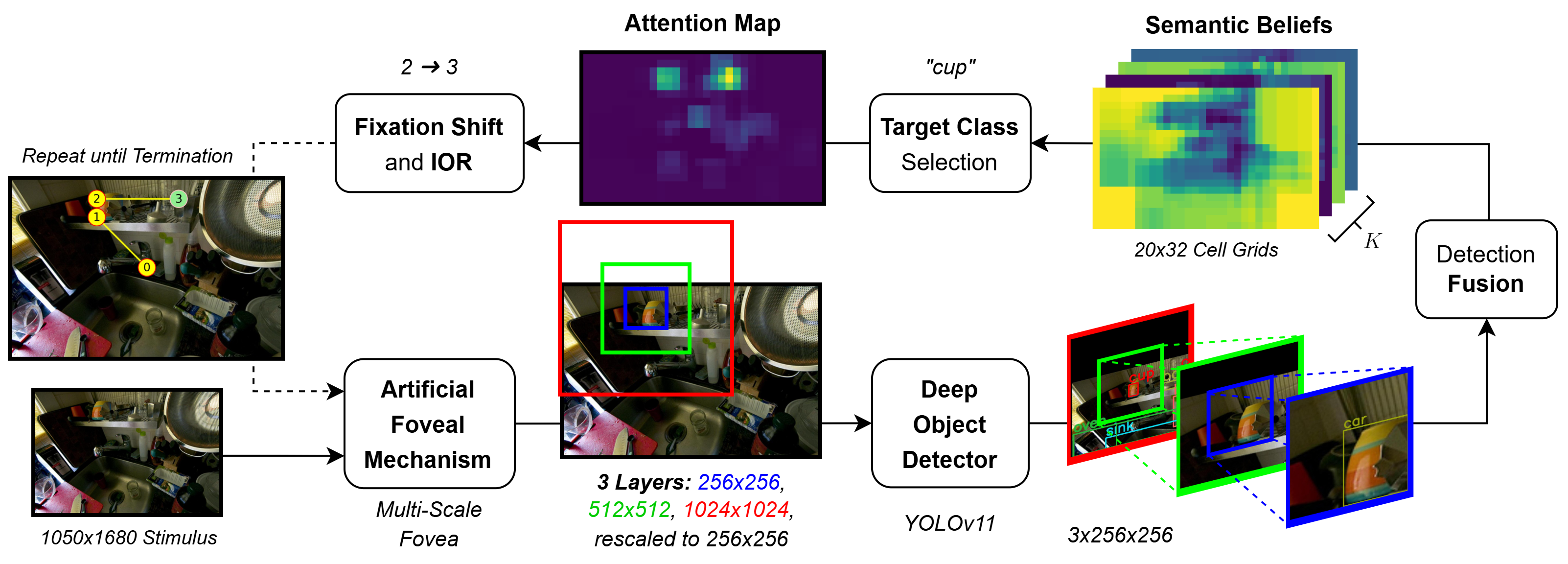}
\caption{Schematic representation of the Semantic-based Bayesian Attention framework, i.e. \textit{SemBA} \cite{neurocomp}, for attention prediction, applied to visual search. In this example, we apply the proposed \textit{Multi-Scale Fovea} mechanism, that generates $N=3$ layers around a focal point, starting at a base dimension  of $256\times256$ pixels and doubling in size between consecutive layers. All layers are then down-sampled to the base dimension to emulate peripheral distortion. Note that the artificial foveation block can be replaced by any other mechanism, such as \textit{FOVEA} \cite{fovea} or \textit{Laplacian Foveation} \cite{fovsys}. The same follows for the deep object detection block, where although we apply YOLOv11 \cite{yolov11} any other modern detection model can be integrated, e.g. DETR \cite{detr}, RT-DETR \cite{rtdetr}. Object detections are then filtered and fused on multiple grids of $20\times32$ cells, to build semantic belief maps for all $K$ known classes. \textit{SemBA}'s active perception system then selects the map for a specific targeted class, which is \textit{cup} in this example, and selects the next-best fixation point based on its current belief state. This process is repeated until a termination criterion is met, while applying inhibition of return (IOR \cite{cocosearch18}) to prevent revisiting fixated locations.}
\label{fig:semba}
\end{figure*}

In essence, the behavior of the human active perception mechanism \cite{activevision} hinges on the nature of the visual task, the characteristics of the field-of-view, and the content of the scene. In this work, we address attention prediction in a target-present visual search setting, where the goal is to find a target object that is known to be present on a provided visual \textit{stimulus}. Mainstream attention models operate by localizing and collecting preattentive information \cite{wolfe} to determine the most salient regions and infer plausible human-generated scanpaths. By scanpaths, we refer to sequences of fixation points \cite{predvisualfix,benchmark}.

In recent years, multiple human scanpath prediction models have been proposed: e.g. Gazeformer \cite{gazeformer}, HAT \cite{hat}, CLIPgaze \cite{clipgaze}. Most of these methodologies exploit selective attention mechanisms, present in modern deep learning architectures, to perform feature extraction in a foveated manner. Despite deriving from classical attention methods \cite{trad_sal,activevision}, state-of-the-art models \cite{gazeformer,hat,clipgaze} have begun to incorporate additional data modalities into their learning process and are now trained using eye-tracking data collected from human subjects. However, though dependent on top-down and bottom-up features, our built-in attention system for visual search \cite{gs6} does not rely on external data provided by other human beings. For instance, infants and children don't learn to visually explore their surrounding environment or search for specific toys from scanpaths generated by their caregivers. In this sense, we can assert that visual attention is purely \textit{stimulus} driven \cite{wolfe}. As an alternative to this notable deviation from biologically-inspired attention modeling, we have recently proposed \textit{SemBA} \cite{neurocomp}, a semantic-based probabilistic framework for human scanpath prediction. This purely \textit{stimulus}-driven attention prediction pipeline, illustrated in Fig. \ref{fig:semba}, leverages pre-trained deep object detectors (e.g. YOLO \cite{yolo_og}, DETR \cite{detr}) to extract and fuse top-down semantic cues from multiple fixations. \textit{SemBA} has been shown to be able to perform target-present visual search \cite{icdl}, competing with other state-of-the-art models in terms of human scanpath similarity. However, it is known that input dimensionality constitutes a bottleneck for time and memory costs in object detection \cite{odreview}, hindering \textit{SemBA}'s performance. Nevertheless, due to foveal eccentricity \cite{eccentricity}, our cognitive system cannot process an entire scene at once. Instead, it cumulatively integrates foveal and peripheral knowledge \cite{fov_review} gathered across multiple eye saccades and fixated regions.

In summary, the human foveal system restricts both the amount and the quality of information that can be perceived from a certain viewpoint. This conditioning is reflected on how objects are displayed across different regions of the field-of-view in terms of geometrical appearance and visual acuity. Most modern deep object detection models are pre-trained on large conventional Cartesian image datasets, such as the COCO 2017 dataset \cite{coco}, being able to invariantly detect objects at different resolutions. To exploit this geometric property of detectors we recover the classical foveal visual systems \cite{image_pyramid} and propose a \textit{Multi-Scale Fovea} mechanism (Fig. \ref{fig:cover}) for localized semantic content extraction. Our new foveation module builds a multi-level pyramid by cropping concentric regions of increasing size around a fixation point. All levels are then downsampled to the scale of the innermost layer in order to increasingly degrade the resolution of semantic content, located in the periphery, while preserving each level's geometrical configuration. On the one hand, our \textit{Multi-Scale Fovea} avoids the need to retrain or fine-tune detectors to handle alternative fovea-like input topologies \cite{fovsys,lukanov,fovea}. On the other hand, it promotes an effective computational cost reduction in object detection \cite{od_pyramid}, making semantic-based attention frameworks, e.g. \textit{SemBA} \cite{neurocomp}, more suitable for deployment in bio-inspired real-time systems, such as humanoid robots.
Regarding our work, we highlight the following contributions:
\begin{itemize}

    \item We propose a novel \textit{Multi-Scale Fovea} mechanism for cost-efficient semantic-based visual attention modelling.
    
    \item We incorporate the new fovea module into \textit{SemBA} \cite{neurocomp}, and assess its performance in target-present visual search.
    
    \item We assess human-model scanpath similarity, using off-the-shelf metrics \cite{benchmark}, and compare \textit{SemBA}'s scores and efficiency with state-of-the-art model's results \cite{ivsn,gazeformer,hat,clipgaze}.
    
\end{itemize}
\noindent\textit{Code availability}: An implementation of \textit{SemBA} $\times$ \textit{Multi-Scale Fovea} is available at \textcolor{magenta}{\href{http://github.com/vislab-tecnico-lisboa/SemBA}{github.com/vislab-tecnico-lisboa/SemBA}}.

\section{Background and Related Work}

\subsection{Deep Object Detection}

Object detection is a famous computer vision problem which involves jointly localizing objects via bounding-boxes and assigning them semantic categorical labels. Early methods, largely based on hand-crafted features and multi-stage pipelines, suffered from limited robustness and scalability \cite{odreview}. 

Capitalizing on the hierarchical feature extraction capabilities of convolutional neural networks (CNNs), the YOLO framework \cite{yolo_og} was introduced as a one-stage family of deep detectors designed for real-time object recognition. YOLO approaches detection by treating it as a unified prediction task, where bounding-boxes and class labels are directly inferred in one forward pass. 
By leveraging global image context, earlier YOLO versions achieved faster inference while sacrificing some localization accuracy, especially for small or crowded objects. The YOLO family has evolved over the years \cite{yolo_rev} with new versions (e.g. YOLOv11 \cite{yolov11}) introducing refinements and enhancements to increase the model's performance. 

The remarkable success of transformers in natural language processing \cite{transreview} has prompted growing interest in extending their use to computer vision tasks. The detection transformer (DETR \cite{detr}) is considered the main foundational model for object detection using visual transformers. 
Whereas YOLO mainly relies on local feature processing, exploiting CNNs' convolutional kernels, DETR is able to model spatial and contextual relationships between multiple objects across an entire scene.
However, despite its competitive performance, DETR is slow to converge during the training phase and shows limited effectiveness in detecting small objects.
As a means to reduce computational costs and increase accuracy, recent iterations to the original DETR architecture have been proposed \cite{transreview}, such as DINO, Co-DETR, LW-DETR, and RT-DETR.
In particular, the real-time detection transformer architecture (RT-DETR \cite{rtdetr}) introduces adaptable inference speed, balancing high accuracy with real-time performance.

\subsection{Computational Visual Attention Models}

Human gaze control has been a topic of interest in neurology and psychology \cite{predvisualfix}, given its importance for understanding visual perception and cognition. In recent years, this particular topic has also been gaining prominence among the computer vision community.
The seminal work of L. Itti and C. Koch \cite{trad_sal} paved the way for saliency-based attention, offering an intuitive and neurologically-inspired way to relate bottom-up features (i.e. color contrasts, intensity and orientation) to free-viewing attentional behavior. The Itti-Koch model was the first to exploit the concept of saliency maps, believed to be an integral part of the posterior parietal cortex of early primates.

Despite the advancements on bottom-up saliency models \cite{benchmark}, goal-directed attention was mostly discarded until the rise of artificial neural networks in popularity. The invariant visual search network (IVSN \cite{ivsn}) was a pioneer work on top-down target-present visual search attention, specifically for zero-shot settings. This template matching model processes a generic target image in parallel with the full scene, using two distinct neural networks. An attention map is built by matching the outputs of both networks. To generate scanpaths, likely fixation points are then sequentially selected as the most conspicuous regions. However, the IVSN does not apply any human-like vision system, uniformly processing the entire scene at once.

Similar to object detection, computational attention has developed alongside deep learning \cite{benchmark}, leveraging new architectures and mechanisms to more accurately predict scanpaths.
Modern methods such as Gazeformer \cite{gazeformer} and HAT \cite{hat} exploit the selective attention mechanisms present in vision transformers to predict sequences of fixation points for any provided image. Another model, known as CLIPgaze \cite{clipgaze}, leverages a foundational visual language model (CLIP) pre-trained with large amounts of textual and visual data to predict human-like scanpaths in zero-shot settings. While HAT integrates visual information at two different eccentricities, using peripheral and foveal tokens \cite{hat}, Gazeformer and CLIPgaze do not explicitly account for the characteristics of the human field-of-view. These models have been able to achieve state-of-the-art performance, thriving on the establishment of the first visual search benchmark dataset: COCO-Search18 \cite{cocosearch18}.

COCO-Search18 provides not only the scenes themselves but also the scanpaths of real human subjects for each instance of the dataset. Modern methods \cite{gazeformer,hat,clipgaze} are trained using the visual \textit{stimuli} and their respective human-generated scanpaths. However, as mentioned in Section \ref{intro}, humans don't learn from other humans' fixation data. Our innate attention system is purely guided by information that can only be perceived directly from the field-of-view. Nevertheless, goal-directed attention relies on visual cues that are top-down by nature, such as semantic information. For this reason, we recently proposed a semantic-based Bayesian attention framework (\textit{SemBA} \cite{neurocomp}) for human attention prediction, which leverages pre-trained deep object detectors. In essence, \textit{SemBA} functions as a pipeline for collecting and fusing semantic information (into an attention map) across multiple fixation points. The full implementation of \textit{SemBA} is detailed in Section \ref{semba}.

\subsection{Artificial Foveation Methods}

In the literature, the definition of fovea is often ambiguous, resulting in different assumptions on its actual dimensions. Although the region of maximum acuity, known as foveola \cite{fov}, comprises only a visual angle of \SI{1}{\degree}, intermediate eccentricities between the fovea and the periphery, often referred to as parafovea \cite{fov_review}, can extend roughly from \SI{2}{\degree} to \SI{5}{\degree} in diameter.

To artificially reproduce foveated fields-of-view with such characteristics, computational models \cite{fovea,lukanov,fovsys} apply different types of geometric transformations to regular images. Cesar Bandera and Peter D. Scott's pioneering work on foveal machine visual systems \cite{image_pyramid} set the stage for current research on bio-inspired vision. As an answer to the increasing pixel flow-through rate problem in image processing, their work proposed the usage of linear and exponential pixel density roll-offs in rectangular and hexagonal sampling lattices. Related work \cite{od_pyramid} proved that such technique allows for significant time and memory cost reductions in object detection while maintaining critical geometrical features of Cartesian images. 

Moving away from multi-resolution Cartesian methods and toward more biologically plausible vision, Almeida et al. \cite{fovsys} proposed a Laplacian pyramid that applies radial Gaussian filtering to remove high spatial frequencies on the periphery. This method, which from now on we refer to as \textit{Laplacian Foveation}, has been demonstrated to not interfere with object detection's performance, provided that the non-blurred central region, corresponding to the fovea, covers a large enough area. \textit{Laplacian Foveation} \cite{fovsys} has been already successfully incorporated into the \textit{SemBA} framework \cite{neurocomp} and applied on target-present visual search \cite{icdl} putting on a convincing performance. However, this foveation technique does not promote any effective decrease in terms of computational costs, as object detectors still need to process images in full resolution.

As an attempt to mimic retinal sampling, recent vision models \cite{ozimek,lukanov} apply log-polar transformations on regular images. These more biologically-inspired geometric transforms not only effectively reduce the number of pixels to be processed but are also invariant to centered scaling and rotations. However, accurate object detection requires models to be retrained on large datasets of log-polar-mapped images, which consumes a substantial amount of time and computational resources. Another recent model, known as \textit{FOVEA} \cite{fovea}, uses kernel density estimation from bounding-boxes, generated by an object detector, to construct saliency maps that guide grid-based image warping. High detection density regions are magnified through transforming the original scene into a new warped space, where pixels are sampled according to the estimated saliency. New detections generated in a warped space can then be remapped to the original unwarped scene. 

\begin{figure*}
\includegraphics[width=\linewidth]{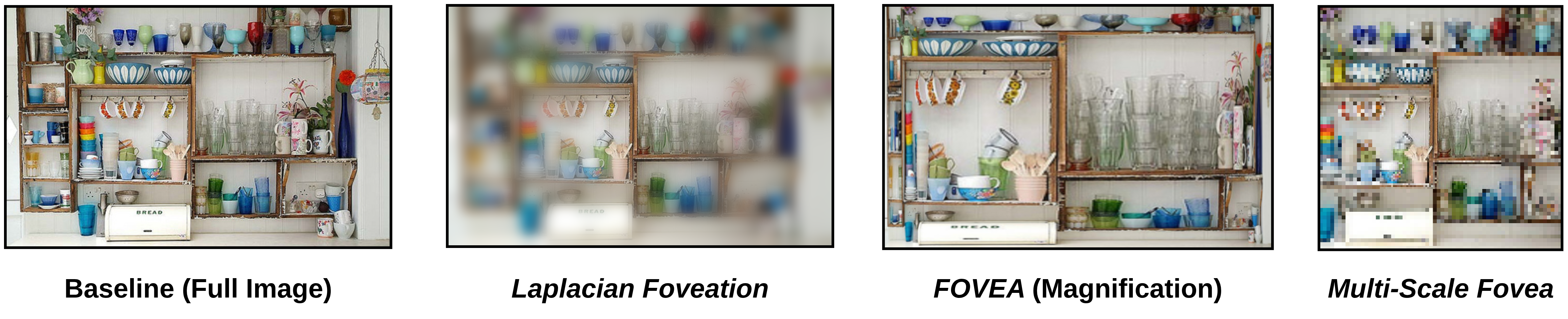}
\caption{Comparison of the field-of-view topologies for each different foveal system used in this work: Full resolution image (baseline), \textit{Laplacian Foveation} \cite{fovsys}, \textit{FOVEA} (Magnification) \cite{fovea}, and our \textit{Multi-Scale Fovea} (in order, from left to right). Here, \textit{Multi-Scale Fovea}'s parameters are set such that the maximum acuity region, corresponding to the fovea \cite{fov}, comprises a region equivalent to around \SI{2.0}{\degree} of visual angle ($64 \times 64$ pixel patch in a $1050 \times 1680$ image). Note that \textit{SemBA} processes each \textit{Multi-Scale Fovea} layer separately. Here, to showcase the topology, we overlap all layers while preserving their resolution.}
\label{fig:fovsystems}
\end{figure*}

\section{Methodology}

In this section, we first outline the preliminary geometric constraints \cite{benchmark} commonly employed in standard human attention prediction setups. We then introduce the probabilistic framework \cite{icdl,neurocomp} adopted for semantic data fusion and next-best fixation prediction in visual search. Finally, we describe our proposed fovea-inspired mechanism for efficient object detection, which reduces computational costs while introducing uncertainty that depends on the visual field's dynamics.

\subsection{Scanpath Prediction Setup}

In line with other state-of-the-art methodologies for human scanpath prediction \cite{ivsn,gazeformer,hat}, we treat two-dimensional images \cite{cocosearch18} as fixed fields-of-view, where the spatial resolution and scene boundaries remain static. We also assume a non-dynamical scene configuration \cite{neurocomp}, where the arrangement of objects within the landscape does not change over time.

Consider an image with dimensions $\textit{height} \times \textit{width}$ and an initial fixation point $f_0$, typically positioned at the center of the visual field. When searching or exploring a given visual \textit{stimulus}, a human produces a sequence of fixation points $f_1, f_2, \dots, f_T$, where each fixation $f_t, \forall t \in \{ 1, \dots, T\},$ corresponds to a specific pixel location within the image. It should be emphasized that the length of the fixation sequence, $T$, differs across tasks, scenes, and individual subjects \cite{cocosearch18}.

To reduce the action space, we employ a standard two-dimensional Cartesian representation to spatially encode the surrounding environment, structured as a $Y \times X$ context grid. By adopting this representation, we minimize computational complexity while approximating the cognitive system's sensitivity levels, balancing efficiency with biological plausibility.

\subsection{Semantic-based Bayesian Attention Framework} \label{semba}

To assess the impact of different foveal geometries in attention prediction, we must first select an appropriate pipeline for collecting and fusing information. These data, extracted across a sequence of fixation points, are used to cumulatively build attention maps from which next fixation locations are inferred. Hence, we consider our recent framework for semantic-based Bayesian attention, i.e. \textit{SemBA} \cite{neurocomp}, that has already been successfully applied to human scanpath prediction \cite{icdl} in a target-present visual search setting. This method exploits the rich semantic information extracted by modern deep object detectors \cite{odreview} to perform goal-directed fixation prediction.

Start by considering a set $\mathcal{C} \!\subseteq\! \left\{1,\dots,K \right\}$ containing all the classes recognizable by a given object detection model.
An object detection is composed of a score vector $S = \left( s_1, s_2, \thinspace \dots \thinspace,  s_K \right)$, where $s_{k} \geq 0, \forall k \in \mathcal{C}$, and its respective bounding-box $\mathcal{B}$. A set of scores $S$ contains the likelihoods of the presence of an instance of each known class within the limits of $\mathcal{B}$. Notice that $S$ does not need to be normalized.

In a nutshell, \textit{SemBA} maps a scene by updating its current semantic beliefs $\boldsymbol{\beta}$ with likelihoods $S$ from new observations. These observations are sequentially extracted from each fixation $f_t$ along the scanpath. Each belief $\boldsymbol{\beta}^\mathbf{x}$ is associated with a cell $\mathbf{x} = (x,y)$, where $x \in \left\{1,\dots,X \right\}$ and $y \in \left\{1,\dots,Y \right\}$. \textit{SemBA}'s active perception mechanism for visual search aims at shifting the gaze to a cell $\mathbf{x}$ that maximizes a posterior categorical probability $P \left( C^{\mathbf{x}} = k \right)$ for a target class $k$. Dirichlet distributions serve as conjugate prior distributions of both categorical and multinomial distributions. Taking advantage of this property, posteriors are modeled through sets of Dirichlet parameters $\boldsymbol{\beta}^\mathbf{x} \in \mathbb{R}^K$, already introduced as semantic beliefs. To define an initial state of maximum entropy, all beliefs are initialized as $\beta^\mathbf{x}_k = 1, \forall k \!\in\! \left\{ 1, \dots, K \right\}$, conforming to flat Dirichlet distributions that represent non-informative priors. 

Each time a detection's bounding-box $\mathcal{B}$ overlaps the region that corresponds to a cell $\mathbf{x}$, the beliefs of that cell are updated:
\begin{equation}
    \beta_{k}^{\mathbf{x}} \longleftarrow \dfrac{\beta_{k}^{\mathbf{x}} \left( 1 + \dfrac{s_{k}}{\sum_{j=1}^{K} \beta_{j}^{\mathbf{x}} s_{j}} \right)}{1 + \dfrac{\min_i s_{i}}{\sum_{j=1}^{K} \beta_{j}^{\mathbf{x}} s_{j}}} , \forall k \in \mathcal{C},
    \label{eq:kaplanrule}
\end{equation}
with $s_k \in S$ representing that detection's categorical likelihood for each class $k \in \mathcal{C}$. This update rule, borrowed from Kaplan's work \cite{kaplan} on classifier fusion, applies subjective logic to incorporate and manage uncertainty in the scores $S$.

\noindent Finally, to determine the next-best viewpoint $\mathbf{x}^\ast$ at time $t$, \textit{SemBA} greedily selects the cell $\mathbf{x}$ with maximum expectancy
\begin{equation}
    \mathbf{x}^\ast = \argmax\limits_{\mathbf{x}} \thinspace \mathbb{E} \left[ C^{\mathbf{x}} = k \mid \boldsymbol{\beta}^\mathbf{x} \right],
    \label{eq:gazeselect}
\end{equation}
where $\mathbb{E} \left[ C^{\mathbf{x}} = k \mid \boldsymbol{\beta}^\mathbf{x} \right] = \beta_k^\mathbf{x} / \sum_{j=1}^{K} \beta_{j}^{\mathbf{x}}$. The gaze is then shifted to the next-best viewpoint $f_{t+1} \equiv \mathbf{x}^\ast$ (in pixels), according to the information gathered from $f_{0:t}$ and fused in $\boldsymbol{\beta}^\mathbf{x}$. \textit{SemBA} iteratively repeats this process, of collecting and fusing new detections from each new $f_{t+1}$, until some terminal criterion is met, while applying inhibition of return (IOR \cite{cocosearch18}).

\subsection{Multi-Scale Fovea Module}

Drawing inspiration from the seminal work of Bandera and Scott \cite{image_pyramid}, we now describe the implementation of our foveal mechanism for efficient bio-inspired semantic data extraction.

Let us start by considering a total of $N$ layers $L_1, \dots, L_{N}$. Each layer $L_n$ consists of a square-shaped crop, centered on a focal point $f \in \mathbb{N}^2$, with sides of length $l_n \in \mathbb{N}$ (in pixels). Hence, a level $L_n$ can be represented by the coordinates of its top-left ($-$)
and bottom-right ($+$) coordinates in the main image frame, i.e. $L_n = \left[ c^{-}_n, c^{+}_n \right]$ such that $c^{\pm}_n \in \mathbb{N}^2$, similar to a bounding-box. Therefore, the corner coordinates of each layer $L_n$ depend on $f=(x_c,y_c)$ and $l_n$ as: $c^{\pm}_n = \left( x_c \pm l_n / 2,\thinspace y_c \pm l_n / 2 \right)$.
The pseudo-foveal region is assumed to be contained within a base layer $L_1$ with an assigned side length $l_1$. To allow the fovea to move close to the scene's boundaries, we apply zero padding to the original image. The side lengths for subsequent layers are then defined as $l_{n+1} = 2 l_{n}$, such that $l_n = 2^{n-1} l_1$, $\forall n \in \{ 1, \dots, N\}$. After being cropped, all layers are then downsampled to the size of the layer $L_1$ via bilinear interpolation. However, $L_1$ preserves its original dimensions, retaining maximum acuity, just as the human fovea. This process, known as exponential pixel density roll-off \cite{image_pyramid}, more intensely degrades the visibility of objects rendered in the outermost layers, akin to peripheral distortion.

All layers $L_n$ are then fed to a deep object detection architecture \cite{yolov11,detr,rtdetr} 
to extract their semantic content. Detections gathered from each layer must first be remapped to the original image $Y \times X$ grid and only then appropriately fused \eqref{eq:kaplanrule} using Kaplan's rule. Let us consider an arbitrary detection from a layer $L_n$, with its corresponding bounding box $\mathcal{B}^{\prime} = \left[ p^{\prime}_{-}, p^{\prime}_{+} \right]$, represented by its coordinates in the layer frame: $p^{\prime}_{\pm} = \left( x^{\prime}_{\pm}\, y^{\prime}_{\pm} \right)$. Finally, we transform the coordinates of $\mathcal{B}^{\prime}$ to the main image frame, and obtain $\mathcal{B} = \left[ p_{-}, p_{+} \right]$ as
\begin{equation}
    p_{\pm} = \left( x_c - \dfrac{l_n}{2} + x^{\prime}_{\pm} 2^{n-1}, y_c - \dfrac{l_n}{2} + y^{\prime}_{\pm} 2^{n-1}\right).
    \label{eq:box_remapping}
\end{equation}
Kaplan's rule \eqref{eq:kaplanrule} is applied to each detection obtained from every single layer, such that multiple observations, potentially corresponding to the same object, are fused sequentially within a single fixation $f_t$. Because this fusion rule promotes low-variance estimates between consecutive updates \cite{kaplan}, aggregating a larger number of high-confidence observations at a given location results in increased posterior confidence for the associated category. Objects nearer to the fovea’s center are processed across a greater number of scales, yielding more corresponding observations. As expected, this leads to more confident estimates for objects displayed closer to the fovea and more uncertain estimates for objects located further into the periphery, mirroring the human visual-cognitive system. 

Since $N\times l_1\times l_1 << \textit{height}\times\textit{width}$, we can assert that this foveal mechanism effectively reduces the total amount of pixels to be processed by any given detector, provided that the base layer size is small enough, i.e. $l_1 < \min\{\textit{height},\textit{width}\}$.

\section{Experiments and Results}




\begin{figure*}
    \centering
    \subfigure[Deep Object Detection Models]{\includegraphics[width=0.32\linewidth]{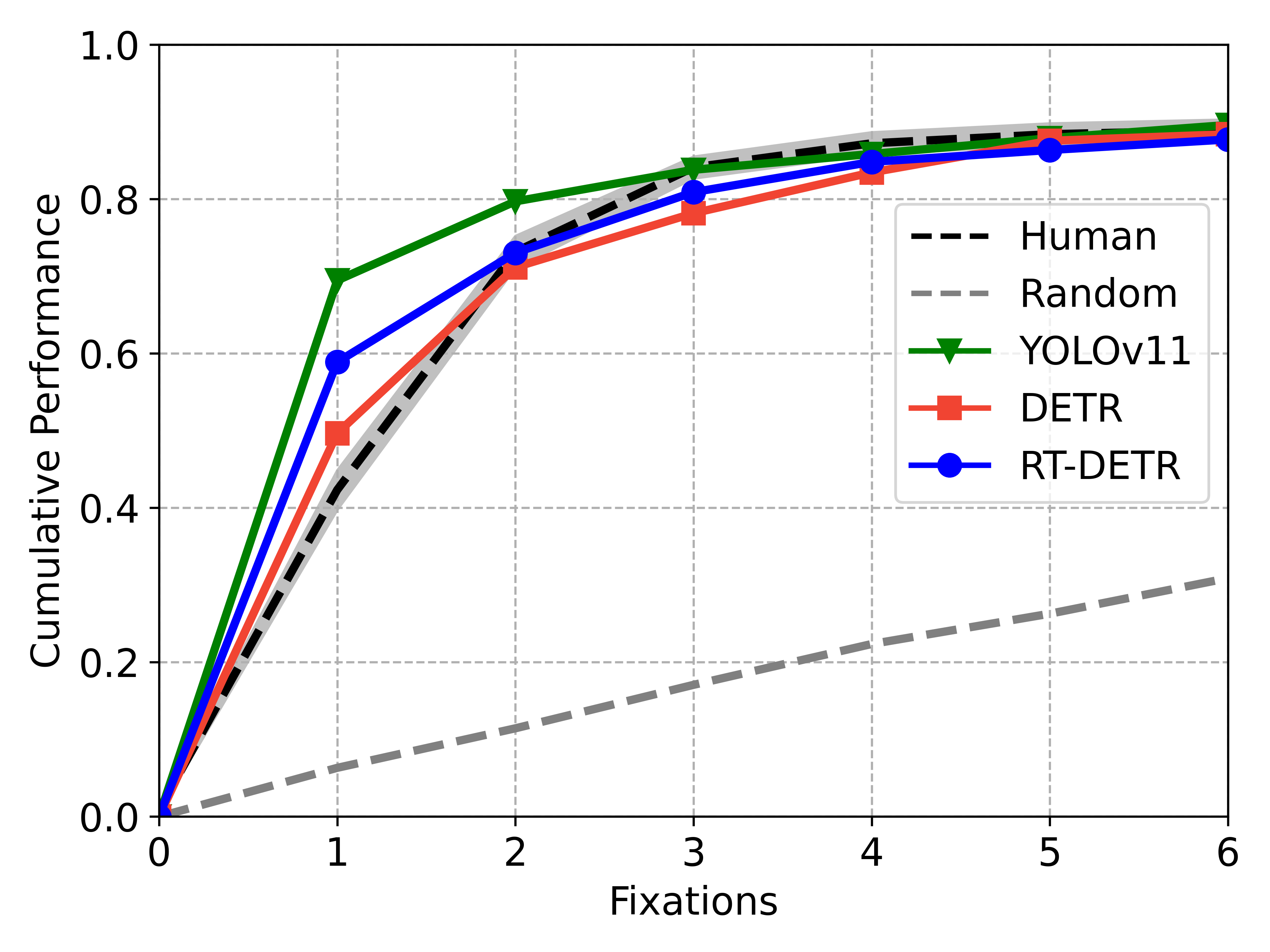}}
    \subfigure[Artificial Foveal Systems]{\includegraphics[width=0.32\linewidth]{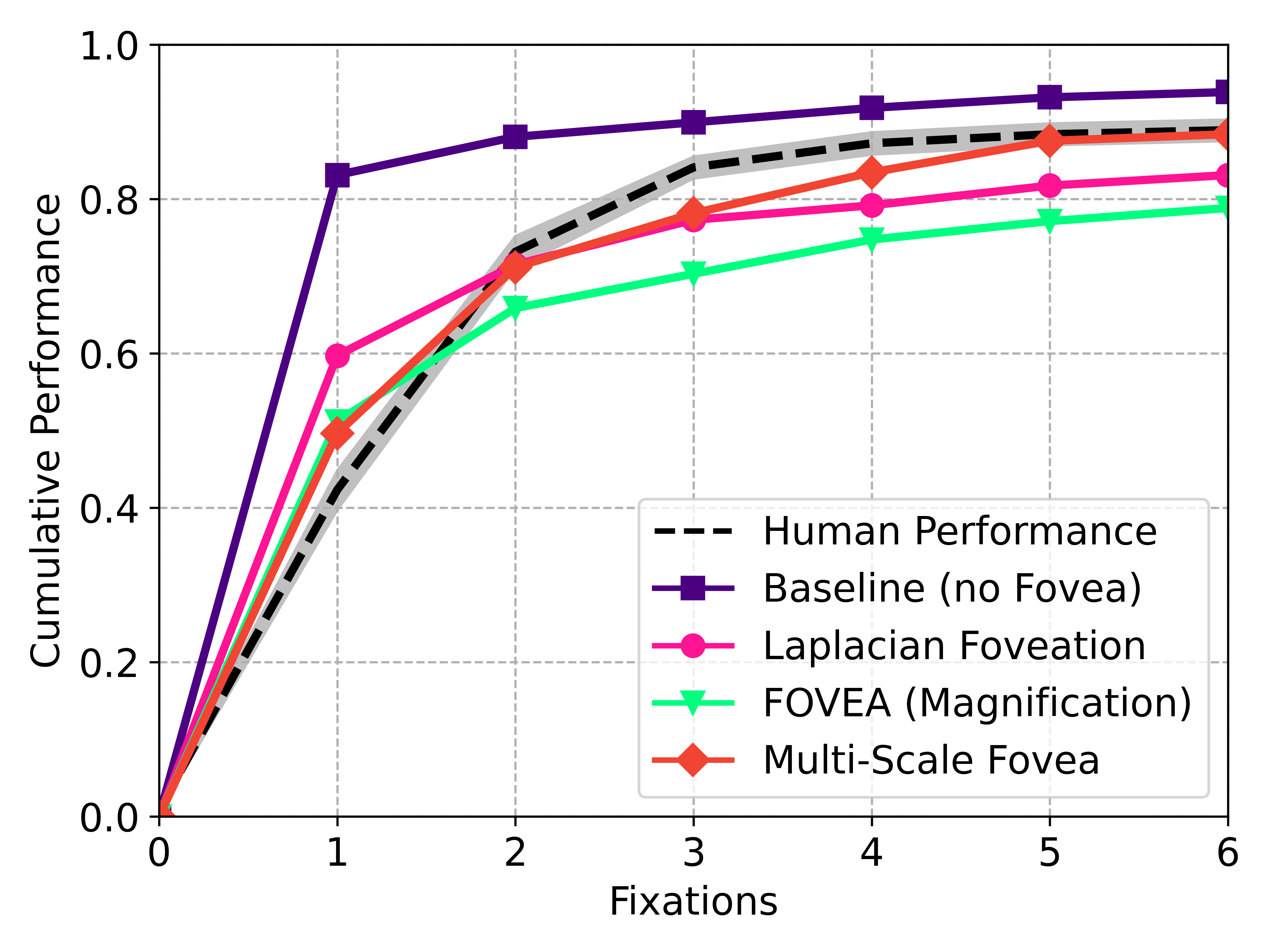}} 
    \subfigure[\textit{Multi-Scale Fovea Settings}]{\includegraphics[width=0.32\linewidth]{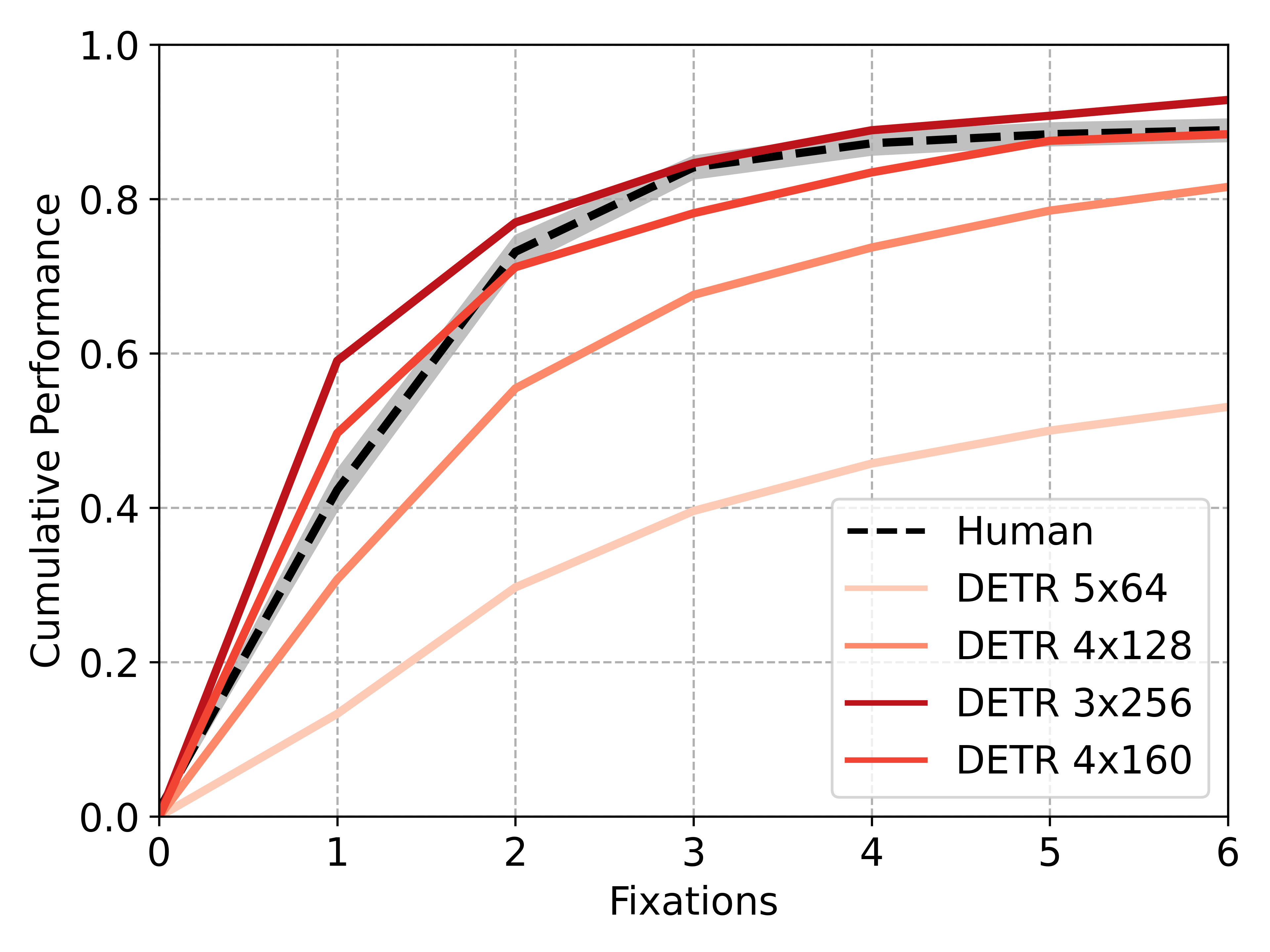}} 
    \caption{Cumulative performances of humans (average), a random selection algorithm, and \textit{SemBA} under different (a) object detectors, (b) foveal systems, and (c) \textit{Multi-Scale Fovea} configurations, on target-present visual search. Regarding experiment (a) we apply our \textit{Multi-Scale Fovea} on a $4 \times 160$ configuration. We use the same configuration for experiment (b) when assessing the performance of the \textit{Multi-Scale Fovea}. For the experiments (b) and (c) we use \textit{SemBA}$\times$DETR.}
    \label{fig:cumulative}
\end{figure*}

\subsection{Experimental Setup}

To appropriately assess the performance of the proposed \textit{Multi-Scale Fovea} (incorporated into \textit{SemBA} \cite{neurocomp}) in target-present visual search, we use off-the-shelf sequence matching metrics \cite{gazeformer,hat,clipgaze}. To comparatively evaluate the impact of the proposed foveal system, we will consider two other types of foveated fields-of-view: \textit{Laplacian Foveation} \cite{fovsys} and \textit{FOVEA} \cite{fovea}. Moreover, given that \textit{SemBA} heavily relies on efficient and accurate object recognition \cite{icdl}, we conduct ablation studies using three distinct state-of-the-art object detection models: DETR \cite{detr}, YOLOv11 \cite{yolov11}, and RT-DETR \cite{rtdetr}.

COCO-Search18 \cite{cocosearch18} is a subset of the well-known COCO 2017 \cite{coco} dataset, comprising a total of 6202 images, 3101 target-present (TP) and 3101 target-absent (TA) scenes, each associated with one of the 18 possible target object categories. For each instance, whether TP or TA, there were collected the natural scanpaths of 10 human subjects, using modern eye-tracking technology. Besides the TP/TA setting distinction, the dataset is split into 3 partitions: train, validation and test sets. To promote a fair comparison with other scanpath prediction models, we evaluate each method on COCO-Search18's test partition alone. COCO-Search18's test set comprises COCO 2017 train and validation samples. Because most deep object detection models \cite{yolov11,detr,rtdetr} are pre-trained on the full COCO 2017 dataset, we are forced to retrain the models, without the samples that are part of COCO-Search18, to prevent information leakage. Therefore, we retrain YOLOv11, DETR, and RT-DETR for 300 epochs using default hyperparameters.

The models are trained to localize and classify $80$ distinct object categories. To ensure that objects located on the blurry peripheral region of the field-of-view are also detected, we set the confidence threshold to \SI{1}{\percent} for all models. By fusing such ambiguous semantic scores we intentionally add uncertainty to the attention mapping process. DETR's architecture \cite{detr} uses ResNet-50 as backbone, adding dilation and removing a stride from the first convolution of its last stage. We also consider large YOLOv11 and RT-DETR (HGNetv2) implementations \cite{yolov11}, containing 25.3 and 32 million parameters, respectively.

As illustrated in Fig.~\ref{fig:semba}, \textit{SemBA} splits an arbitrary size image into a $20 \times 32$ grid. A fixation point is defined as the center-most pixel of a selected cell. After each fixation, IOR is applied in a $3 \times 3$ block \cite{cocosearch18}, centered on the cell that was being fixated. This corresponds to a visual angle of \SI{5.0}{\degree} in diameter, which is similar to the area covered by the parafovea \cite{fov_review}. When dividing a $1050\times1680$ COCO-Search18 image into a $20 \times 32$ grid, a $3 \times 3$ block of cells covers roughly a $160 \times 160$ pixel patch. Hence, when comparing with other scanpath prediction models (Tab. \ref{tab:scanpath_eval}), \textit{SemBA}'s fovea module builds a 4 level pyramid with base length $l_1 = 160$. To assess fovea's size impact on search accuracy and human scanpath similarity (Tab.~\ref{tab:fovcost_eval}) we experiment with \textit{SemBA} and \textit{Multi-Scale Fovea} assigning larger ($l_1 = 256$) and smaller ($l_1=64$ and $l_1 = 128$) dimensions to the layers. The number of levels is chosen so that each configuration encapsulates the exact same region: $N=3$ for $l_1 = 256$, $N=4$ for $l_1 = 128$, and $N=5$ for $l_1 = 64$, covering a total area of $1024 \times 1024$ pixels. Such settings ensure that the majority of semantic information is available (i.e. displayed in the field-of-view) when fixating on the center of any $1050 \times 1680$ image. Both \textit{Laplacian Foveation}~\cite{fovsys} and \textit{FOVEA}~\cite{fovea} topologies follow the configurations illustrated in Fig.~\ref{fig:fovsystems}. Due to its high sensitivity to the choice of foveal proportions, we set the radius of the \textit{Laplacian Foveation} to be the diagonal length of a $3 \times 3$ block of cells. This choice makes the effective foveal area comparable to that of a $3 \times 256$ \textit{Multi-Scale Fovea} configuration. Regarding \textit{FOVEA}, we apply a low-variance Gaussian distribution, centered around the focal point, to generate the warped field-of-view, which compresses the visual appearance of objects located in the peripheral region.

In each experiment, we generate scanpaths for 586 samples from COCO-Search18's test set, in their maximum resolution ($1050\times1680$). By benchmark convention \cite{cocosearch18}, search is always initiated ($f_0$) at the center of each test image. Similar to IVSN \cite{ivsn}, we apply an oracle that stops the search process once the gaze is set upon the targeted object's ground-truth bounding-box. By opting for an oracle, we avoid the need to tackle the challenging problem of defining a hard confidence threshold $\tau$ that terminates the search process, i.e. $\mathbb{E} \left[ C^{\mathbf{x}} = k \mid \boldsymbol{\beta}^{\mathbf{x}} \right] \geq \tau$. 

\subsection{Evaluation Metrics}

We now describe the off-the-shelf benchmark metrics \cite{hat} used for assessing the similarity between ground-truth human scanpaths and model-generated fixation sequences. Sequence Score (\textbf{SS}) represents scanpaths as ordered strings of fixation cluster IDs and quantifies their similarity through the Needleman–Wunsch global alignment algorithm \cite{needlewunsch}. This algorithm was originally designed for comparing between two protein amino-acid sequences. In contrast, Fixation Edit Distance (\textbf{FED}) encodes scanpaths using the same cluster-based string representation but evaluates dissimilarity via Levenshtein’s edit distance \cite{levenshtein}. Semantic Sequence Score (\textbf{SemSS}) departs from SS by replacing spatial cluster identifiers with categorical semantic labels that denote the objects fixated at each step, while retaining Needleman–Wunsch's algorithm. Analogously, Semantic Fixation Edit Distance (\textbf{SemFED}) operates on semantic label sequences but adopts Levenshtein’s edit distance.

To quantify human consistency, as reported in Tab. \ref{tab:scanpath_eval} and Tab. \ref{tab:fovcost_eval}, we compute the four distance metrics (SS, FED, SemSS, SemFED) between all pairs of scanpaths (from 10 human subjects) for each COCO-Search18 test image. For each considered trial, we ignore whether the subject's response was correct or incorrect. Table entries correspond to the mean value of each metric, averaged over all possible scanpath pairs. Moreover, to assess task performance, we also plot the cumulative performance (Fig.~\ref{fig:cumulative}) attained by each model or foveal setting. It consists of the ratio between the number of samples where the target was already found and the total number of test samples, after a certain number of fixations. 

To assess model performance \cite{benchmark}, we compare the scanpath generated by the model with the scanpaths generated by human subjects for the same test sample, averaging the respective metric scores. To promote a fair comparison \cite{icdl}, we truncate the lengths of all scanpaths to 6 fixations (excluding $f_0$). For each metric, we highlight the best model’s value in bold, while underlining other models’ values that beat human consistency.

\begin{table}
\centering
\caption{Scanpath Prediction Model Evaluation in Visual Search}
\begin{tabular}{c|c|c|c|c} \toprule
                  & SemSS $\uparrow$ & SemFED $\downarrow$ & SS $\uparrow$ & FED $\downarrow$ \\ \midrule
Human Consistency & 0.470 & 2.144 & 0.463 & 2.353 \\ \midrule
IVSN \cite{ivsn}  & 0.380 & 3.034 & 0.346 & 3.360 \\
Gazeformer \cite{gazeformer}      & \underline{0.490} & \underline{1.928} & \textbf{0.504} & \underline{2.072} \\
HAT \cite{hat}    & \underline{0.540} & \underline{1.522} & \underline{0.468} & \underline{2.063} \\
CLIPgaze \cite{clipgaze}          & \textbf{0.545} & \textbf{1.489} & \underline{0.476} & \textbf{2.014} \\ \midrule
\textit{SemBA} $\times$ YOLOv11$^{\ast}$  & 0.448 & 2.161 & 0.414 & 2.572 \\
\textit{SemBA} $\times$ DETR$^{\ast}$     & \underline{0.488} & \underline{2.102} & 0.426 & 2.574 \\ 
\textit{SemBA} $\times$ RT-DETR$^{\ast}$  & \underline{0.473} & 2.167 & 0.421 & 2.635 \\ \bottomrule
\multicolumn{5}{l}{$^{\ast}$Incorporating our \textit{Multi-Scale Fovea} in a $4\times160$ pixel configuration.}
\end{tabular}
\label{tab:scanpath_eval}
\end{table}

\subsection{Results and Discussion}

In Fig.~\ref{fig:cumulative} we show the cumulative performance of \textit{SemBA} in target-present visual under different settings. For comparison, we present the results of a random gaze selection method as well as the average human performance (across 10 subjects) together with the respective standard error of the mean bands. 

Setting our \textit{Multi-Scale Fovea} parameters to cover a \SI{5.0}{\degree} angle ($4 \times 160$ configuration), we assess our attention pipeline's performance when incorporating different (a) object detection and (b) foveal mechanism modules. From the curve presented in Fig.~\ref{fig:cumulative} (a) we observe that, under the mentioned fovea setup, \textit{SemBA} achieves the human average cumulative performance after 6 fixations (around \SI{90}{\percent}), whether paired with YOLOv11, DETR, or RT-DETR. However, regarding the performance after just 1 fixation, YOLOv11, RT-DETR, and DETR (about \SI{70}{\percent}, \SI{60}{\percent}, and \SI{50}{\percent}, respectively) tend to overshoot human results (near \SI{40}{\percent}). Nevertheless, DETR's performance curve is clearly the most similar to the human curve. Therefore, we select DETR as our go-to model when assessing the artificial fovea module configurations in Fig.~\ref{fig:cumulative}. 

Regarding scanpath similarity, Tab.~\ref{tab:scanpath_eval} reveals that \textit{SemBA} is able to level human consistency in terms of sequence semantic similarity (SemSS and SemFED), even slightly surpassing it. However, \textit{SemBA} is still below human consistency in terms of sequence location distance (SS and FED). On the one hand, given that \textit{SemBA} relies solely on semantic information, it is critical to conclude that produced sequences of fixated objects (both target and distractors) accurately mimic human patterns. On the other hand, the fact that \textit{SemBA} does not learn from actual human scanpaths possibly hinders its fixation location accuracy when compared to other benchmark models \cite{gazeformer,hat,clipgaze}. However, with the new \textit{Multi-Scale Fovea}, \textit{SemBA} still outperforms \cite{icdl} the best performing scanpath prediction model that does not learn from human scanpaths, i.e. IVSN \cite{ivsn}. When analyzing semantic distance metrics, \textit{SemBA} is the model that best approximates human consistency, despite underperforming compared to state-of-the-art models \cite{gazeformer,hat,clipgaze}, which substantially overshoot inter-human performance.

\begin{table}
\centering
\caption{Ablation Study and Computational Cost Evaluation}
\begin{tabular}{c|c|c|c} \toprule
                  & SemSS$^{\ast}$ $\uparrow$ & SemFED$^{\ast}$ $\downarrow$ & Pixels $\downarrow$\\ \midrule
    Human Consistency   & 0.470 & 2.144 &  -  \\ \midrule
    Baseline (no Fovea) & 0.430 & \textbf{2.025} & \SI{100}{\percent} \\
    \textit{Laplacian Foveation} \cite{fovsys} & 0.416 & 2.488 & \SI{100}{\percent}  \\
    \textit{FOVEA} (Magnification) \cite{fovea}           & 0.384 & 2.769 & \SI{100}{\percent}  \\ \midrule
    \textit{Multi-Scale Fovea} $5\times64$   & 0.351 & 3.771 & \SI{0.01}{\percent}  \\
    \textit{Multi-Scale Fovea} $4\times128$  & 0.460 & 2.409 & \SI{3.72}{\percent}  \\
    \textit{Multi-Scale Fovea} $3\times256$  & 0.464 & \underline{2.099} & \SI{11.1}{\percent} \\ 
    \textit{Multi-Scale Fovea} $4\times160$  & \textbf{0.488} & \underline{2.102} & \SI{5.80}{\percent} \\\bottomrule
    \multicolumn{4}{l}{$^{\ast}$Original image resolution of $1050\times1680$, applying  \textit{SemBA}$\times$DETR.}
    \end{tabular}
\label{tab:fovcost_eval}
\end{table}

With respect to the evaluation of the artificial foveation module's impact on \textit{SemBA} performance, using DETR, Fig.~\ref{fig:cumulative} (b) shows that, for the selected parameters, the \textit{Multi-Scale Fovea} obtains the best task performance after 6 fixations. Furthermore, its curve better approximates the human curve when compared to the other foveal topology curves. Regarding semantic distance metrics, Tab.~\ref{tab:fovcost_eval} reveals that only the newly proposed fovea is able to surpass human consistency, beating both \textit{Laplacian Foveation} and \textit{FOVEA}. As highlighted, the key advantage of the \textit{Multi-Scale Fovea} is that it heavily reduces the amount of pixels to be processed by the object detector.
Note that \textit{FOVEA} (and possibly \textit{Laplacian Foveation}) could eventually yield more accurate results if the selected object detector was retrained after applying the respective foveal transforms to the training dataset. However, because our \textit{Multi-Scale Fovea} preserves critical geometric properties of regular images, we avoid the need to retrain object detectors, which is a very time-consuming and computationally costly process.

Finally, we conduct an ablation study to determine the influence of the \textit{Multi-Scale Fovea} parameters. As expected, Fig.~\ref{fig:cumulative} (c) conveys that decreasing the fovea dimensions generates a lower cumulative performance due to the increased peripheral distortion, and vice versa. As previously mentioned the $4\times160$ configuration, which is more congruent with the actual fovea anatomical proportions, better approximates the human curve while the $3\times256$ setting essentially serves as its upper-bound. Moreover, unlike in the other configurations, only in the $4\times160$ setting does \textit{SemBA} excel human consistency across both semantic sequence similarity metrics, as shown in Tab.~\ref{tab:fovcost_eval}.

To assess whether our novel \textit{Multi-Scale Fovea} effectively promotes computational gains in object detection, we run inference (NVIDIA GeForce RTX 4060) on a full $1050 \times 1680$ image and a $4 \times 160$ setting, using YOLOv11 and DETR. In terms of time cost (in seconds), \textit{SemBA} $\times$ DETR drops from \SI{10.37}{\second} to \SI{0.59}{\second} per iteration, achieving a notable $17.6$x speed-up. However, \textit{SemBA} $\times$ YOLOv11 only drops from \SI{0.144}{\second} to \SI{0.115}{\second}, showing a minor speed-up. This results show that the proposed \textit{Multi-Scale Fovea} considerably diminishes detection costs for heavier models (e.g. DETR), but promotes minimal gains when applying modern architecture (e.g. YOLOv11) that are by now prepared to swiftly handle larger visual inputs. Nevertheless, modern architectures, which generally achieve higher levels of accuracy, appear to inadequately capture the uncertainty 
that is inherently linked to the attentional behavior.


\section{Conclusions}

Deep object detection models excel at extracting top-down semantic cues \cite{odreview}, which constitute one of the main sources of preattentive information. 
Because the size of the visual input has been traditionally considered a bottleneck in object detection \cite{od_pyramid}, we propose a \textit{Multi-Scale Fovea} mechanism that reduces the total amount of pixels to be processed from each fixated point. Inspired by the anatomy of the human visual system, our novel method builds a multi-resolution pyramid \cite{image_pyramid}, around a focal point, which gradually degrades the quality of information in more peripheral levels. We show that the \textit{Multi-Scale Fovea} is able to improve the performance \cite{icdl} of our semantic-based Bayesian attention framework (\textit{SemBA}) in target-present visual search. The proposed hard-attention mechanism leads \textit{SemBA} to more closely mimic human gaze patterns, notably in terms of the sequences of fixated object categories. Therefore, we show that the relationship between overt and covert attention can be effectively modeled within an efficient framework that learns solely from cues extracted directly from visual \textit{stimuli}.
As future work, we aim to extend the proposed framework and foveal system to other visual tasks, namely free-viewing and visual question answering.

\end{document}